\documentstyle[prd,twocolumn,aps,tighten,floats]{revtex}
\begin{document}
\draft
\title{5D Kaluza - Klein's Wormhole Between Two Event Horizons}
\author{Dzhunushaliev V.D.
\thanks{E-mail address: dzhun@freenet.bishkek.su}}
\address {Theoretical Physics Department \\
          Kyrgyz State National University, Bishkek, 720024}
\maketitle
\begin{abstract}
{The solution of the 5D Kaluza-Klein's theory is obtained.
This solution is a Lorentzian wormhole between two event 
horizons. It is shown that this solution can sew with (4D
Reissner-Nordstr\"om's solution $+$ Maxwell's electrical
field) on the event horizon. Such construction is composite
wormhole connecting two asymptotically flat region. From the 
viewpoint of infinite observer this wormhole is an electrical 
charge. According to J.Wheeler terminology this is "charge 
without charge".
}
\end{abstract}

In contemporary models of quantum gravity it is of prime 
important the dynamical process carrying out change of the 
space - time topology \cite{wh}. It is supposed that such processes 
take place as a result
of the wormholes origin. There are many models of the Eucledean's 
wormholes filled some kind of matter field (see, for example, 
\cite{re}, \cite{jo}).
\par
The aim of this paper is the construction of a composite wormhole:
it is the solution of 5D Einstein's equations (Kaluza - Klein's equation)
under event horizon (EH), and Reissner - Nordstr\"om's solution
outside EH. Onto EH is performed sewing the 5D Kaluza - Klein's 
metric with 4D Einstein's metric + electrical field.

\subsection{5D Wormhole in Kaluza - Klein's theory}

The initial 5D Kaluza - Klein's equation's write down in the 
following standart manner:
\begin{equation}
R_{\mu\nu} - {1\over 2} G_{\mu\nu} R = 0
\label{11}
\end{equation}
where $G_{\mu ,\nu}$ is 5D metrical tensor; $R_{\mu\nu}$ is Ricci
tensor; $R$ is scalar curvature; $\mu ,\nu = 1,2,3,4,5$. 5D metric we 
seek in the following wormholelike view:
\begin{equation}
ds^2 = e^{2\nu (r)}dt^2 - e^{2\psi (r)} \left (d\chi - \omega (r)dt
\right )^2 - dr^2 - a^2(r)\left (d\theta ^2 + \sin ^2\theta d\varphi ^2
\right ),
\label{12}
\end{equation}
where $\chi$ is 5 supplementary coordinate; $r, \theta ,\varphi$ are
3D polar coordinates; $t$ is the time. Corresponding Kaluza - Klein's 
equations have the following view:
\begin{eqnarray}
\psi '' + {\psi '}^2 + \psi '\nu ' + {{2a'\psi '}\over a} +
{{q^2}\over {2a^4}} e^{-4\psi} = 0,
\label{131}\\
a'' + a'\psi ' + {{{a'}^2}\over {a}} + a'\nu ' - {1\over a} = 0,
\label{132}\\
\nu '' + {\nu '}^2 + \psi '\nu ' + {{2a'\nu '}\over a}-
{{q^2}\over {2a^4}} e^{-4\psi} = 0,
\label{133}\\
\psi '' + {{2a''}\over{a}} + {\psi '}^2 - \psi '\nu ' - 
{{2a'\nu '}\over a} = 0,
\label{134}
\end{eqnarray}
here $(')$ denotes the derivative with respect to $r$. In equations 
system $(\ref{131}-\ref{134})$ it is taken the following 
equation into account:
\begin{equation}
\omega ' = {{q}\over{a^2}}e^{-3\psi + \nu},
\label{14}
\end{equation}
obtained by integration of $R_{\chi t} = 0$ equation. $q$ 
in Eq.(\ref{14}) is some constant ("electrical" Kaluza - Klein's charge).
\par
From Eq's $(\ref{131})$ and $\ref{133}$ it follows that $\psi = -\nu$
and the initial system $(\ref{131}-\ref{134})$ can written
as follows:
\begin{eqnarray}
a'' + {{{a'}^2}\over{a}} - {{1}\over{a}} = 0,
\label{151}\\
\nu '' + {{2a'\nu '}\over a} - {{q^2}\over{2a^4}}e^{4\nu} = 0,
\label{152}\\
{\nu '}^2 + {1 \over{a^2}}\left (1 - {{q^2}\over{4a^2}} e^{4\nu}\right )
-{{{a'}^2}\over{a^2}} = 0.
\label{153}
\end{eqnarray}
\par
The integration of the Eq.$(\ref{151})$ and Eq.$(\ref{153})$ leads to
the following result:
\begin{eqnarray}
a^2 = r^2_0 + r^2,
\label{161}\\
e^{2\nu} = {{2r_0}\over q} {{r^2_0 + r^2}\over{r^2_0 - r^2}}.
\label{162}
\end{eqnarray}
where $r_0$ is a minimal radius of given wormhole. Eq.$(\ref{152})$
is identical satisfied.
\par
The integration of Eq.$(\ref{14})$ leads to the following result:
\begin{equation}
\omega = {{4r^2_0}\over q} {r\over{r^2_0 - r^2}}.
\label{17}
\end{equation}
\par 
It can be shown that the time component of metrical tensor
$G_{tt}(r=\pm r_0)=0$. This indicates that there is the event 
horizon by $r=\pm r_0$.
\par
It is easy to see that solution $(\ref{161}-\ref{17})$ is nonsingular
by $r\in(-\infty, +\infty)$ but the supplementary coordinate $(\chi)$
can turns out to the timelike coordinate by $|r|>r_0$. This is 
unsatisfactory with respect
to physical point of view. On the other hand if the Kaluza - Klein's
action:
\begin{equation}
S = -\int R \sqrt{-G} d^5x
\label{a}
\end{equation}
not vary with respect to $G_{tt}$ component, then we obtain the regular 
physically reasonable solution by $|r|>r_0$. This is 
Reissner-Nordstr\"om's solution.

\subsection{The sewing Kaluza - Klein's wormhole with 
Reissner - Nordstr\"om's solution}

In this section we show that the Reissner - Nordstr\"om's solution
really can to sew on EH with above received 5D wormhole \cite{dzh}. 
To do this, the 5D Kaluza - Klein's metrical tensor is necessary 
to sew with (4D
metric + Maxwell's electrical field in Reissner - Nordstr\"om's solution).
\par
The following metric components must be in agreement on the sewing
surface $=EH$:
\begin{eqnarray}
e^{2\nu _0} - \omega ^2_0 e^{-2\nu _0} = G_{tt}\left (r_0\right ) 
= g_{tt}\left (r_+\right ) = 0,
\label{21}\\
r^2_0 = G_{\theta\theta} = g_{\theta\theta} = r^2_+,
\label{22}
\end{eqnarray}
where $G$ and $g$ are 5D and 4D metrical tensor respectively.
$r_+ = m + \sqrt{m^2 + Q^2}$ is EH for Reissner - Nordstr\"om's
solution ($m$ and $Q$ are mass and charge of the Reissner - Nordstr\"om's
black hole). The quantity marked by $(0)$ sign are taken by $r=r_0$.
\par
To sew $G_{\chi t}$ and 4D electrical field we consider 5D 
$R_{\chi t}=0$ equation and Maxwell's equation:
\begin{eqnarray}
\left [a^2\left (\omega 'e^{-4\nu}\right )\right ]' = 0,
\label{23}\\
\left (r^2E_r\right )' = 0,
\label{24}
\end{eqnarray}
here $E_r$ is 4D electrical field.
\par
This 2 equations are practically the Gauss law and they indicate that
some quantity multipled by area is conserved. In 4D case this quantity 
is 4D Maxwell's electrical field and from this follows that the 
electrical charge is 
conserved. Thus, naturally we must join 4D electrical 
Reissner - Nordstr\"om's 
field $E_{RN} = Q/r^2_+$ with "electrical" Kaluza - Klein's field 
$E_{KK} = \omega 'e^{-4\nu}$ on EH:
\begin{equation}
\omega _0'e^{-4\nu _0} = {q\over{2r^2_0}} = E_{KK} = 
E_{RN} = {Q\over{r^2_+}}.
\label{25}
\end{equation}
\par
Now, it is necessary to consider the question on the sewing physical
quantity produced from 1th and 2th derivative of metric. Such geometrical
quantity as connection and curvature tensor consist of the 1th and 2th
derivative of the metrical tensor component respectively. Hence, it is 
necessary to examine the question of sewing this quantity.
\par
By sewing 4D and 5D connections and curvature tensors on EH we obtain a 
singularity (on surface sewing) 
similar to singularity of $|x|$ function at the point $x=0$.
In this point of view 1th derivative has dicsontinuity and 2th 
derivative has 
$\delta$-like singularity. Analogically in our situation the connection has 
discontinuity and curvature tensor has $\delta$-like singularity.
\par
By such singularity the connection as well as the curvature tensor
can to have certain nonsingular geometrical meaning like to gravitational
Regge calculus: The components of the tangent vector, which are transferred
in a parallel way through sewing surface (EH) at an infinitesimal
distance, acquire  the finite increment from the fact that the connection 
has discontinuity.

\end{document}